\documentclass[aps,prb,preprint]{revtex4-1}
\usepackage{graphicx}
\usepackage{amsmath}
\begin{document}

\title{Magnetic states of iron-based superconducting compounds calculated by using GGA$+U$ method with negative \emph{U}}

\author{Jae Kyung Jang}
\author{Joo Yull Rhee}
\email[]{rheejy@skku.edu}
\affiliation{Department of Physics, Sungkyunkwan University, Suwon 440-746, Republic of Korea}

\date{\today}

\begin{abstract}
  The magnetic moments per Fe atom in high-$T{_\textrm{c}}$ iron-based superconducting compounds, BaFe$_{2}$As$_{2}$ and LaFeAsO obtained from the first-principles calculation with local-spin-density approximation are much larger than those obtained from experiments. To resolve the contradictory results between theory and experiment we employed the so-called LDA$+ U$ (or more exactly GGA$+ U$) technique with negative \emph{U} in the first-principles calculation. The calculated values with negative \emph{U}, $- 0.09$ Ry and $- 0.10$ Ry for BaFe$_{2}$As$_{2}$ and LaFeAsO, respectively, are in excellent agreement with the experimental ones. By comparing the differences in \emph{d}-orbital occupation numbers and spin densities calculated by using a simple GGA and GGA$+ U$ with negative $U$, the magnetic moments of the two compounds are found to be similar to the case of low-spin state of metamagnetic Fe$_{3}$Al alloy.
\end{abstract}

\maketitle

\section{Introduction}
The discovery of iron-based superconductors \cite{Kamihara:08} attracted many researchers to devote themselves into searching for possible room-temperature superconductors. Another important aspect of this system is the possibility of coexistence of seemingly mutually exclusive properties; magnetism and superconductivity. Some important issues were resolved, but there are many other aspects whose underlying physical mechanisms are still unsolved. In cases of BaFe$_{2}$As$_{2}$ (referred to it as the 122-compound hereafter) and LaFeAsO (referred to it as the 1111-compound hereafter) compounds, the former undergoes the simultaneous structural, from tetragonal to orthorhombic structures, and magnetic phase transitions, from paramagnetic to antiferromagnetic (AFM) phases, at $\sim$140 K,\cite{Huang:08} while the latter undergoes the structural phase transition from tetragonal to monoclinic (or orthorhombic\cite{Yildirim:08}) structures at $\sim$155 K, and develops long-range spin-density-wave type AFM order at $\sim$137 K.\cite{Cruz:08} The magnetic moment per Fe atom of the 122-compound determined experimentally is about 0.87 $\mu_{\textrm{B}}$,\cite{Huang:08} but the theoretical value is about 2.6 $\mu_{\textrm{B}}$,\cite{Akturk:09} and that of the 1111-compound is about 0.36 $\mu_{\textrm{B}}$ in experiment,\cite{Cruz:08} but is about 1.6 $\mu_{\textrm{B}}$ in theory.\cite{Nomura:08} For both cases, the calculated values are about $3 - 4$ times overestimated over the experimental results.

At the first glance, it is usually tempted to apply the so-called `LDA (local-density approximation) $+ U$' method. This method was developed to properly describe the strongly-correlated-electron (SCE) systems which usually have very localized electrons with high angular momentum. Due to the strong on-site Coulomb repulsion among well-localized electrons, a Hubbard-type onsite repulsion is `manually' included in the exchange-correlation functional during the self-consistent-field (SCF) calculation. This method was very successful for many SCE systems, however, it is well known that the onsite repulsion (i.e, $U > 0$) enhances the magnetism, which is not desirable for our system. Therefore, we performed the first-principle calculations with generalized-gradient approximation (GGA)$+U$ method, especially by using the negative \emph{U}. The negative \emph{U}, more exactly $U_{\textrm{eff}} = U - J$, implies that there is an effectively attractive force among \emph{d}-electrons of Fe, which is very odd, however, there have been some theoretical suggestions already.\cite{Micnas:90,Hase:07,Nakamura:09} Moreover, the material systems, especially superconductors, with a negative effective \emph{U} are not rare.

We found that, in high-$T{_\textrm{c}}$ iron-based superconducting compounds, the use of GGA$+U$ with negative \emph{U} of similar magnitudes ($-0.09$ Ry and $-0.10$ Ry) leads to the well-matched values with experimental magnetic moment per Fe. We demonstrate the validity of the negative \emph{U} by calculating band structures, density of states (DOS), \emph{d}-electron occupation numbers and spin densities of the two compounds. Through these results, we can carefully deduce that the use of negative \emph{U} in the first-principle study can give more accurate results of magnetic moments of iron-pnictide systems. More specifically, the magnetic states of the two compounds are found to be similar to the case of low-spin state of metamagnetic Fe$_{3}$Al alloy.

\section{Theoretical calculations}
We used the WIEN2k package\cite{wien2k} implemented with full-potential linearized-augmented-plane-wave method. The exchange-correlation functional was chosen to be GGA version of Perdew, Burke and Ernzerhof.\cite{Perdew:96} The spin-orbit coupling was not included. We used $RK_{\max}= 7$ and $\sim$ 270 (for the 122-compound) and $\sim$ 220 (for the 1111-compound) augmented plane waves for the basis functions. For SCF cycles we generated 1000 k-points in the whole Brillouin zone (BZ) corresponding to 423 (for the 122-compound) and 424 (for the 1111-compound) $\textrm{\textbf{k}}$-points in the irreducible wedge of the BZ. After self consistency was achieved, we further increase the number of $\textrm{\textbf{k}}$-points to 5000 in the whole BZ to obtain DOS and charge densities.

The AFM and the so-called fixed-spin-moment (FSM) calculations featured in the WIEN2k package were used. It is possible to constrain the total spin magnetic moment per unit cell to a fixed value and thus force a particular ferromagnetic solution (which may not correspond to the equilibrium). This is particularly useful for systems with several metastable magnetic solutions, where conventional spin-polarized calculation would not converge or the solution may depend on the starting density.\cite{wien2k} Since the two compounds are in the AFM state, the total magnetic moment are fixed to be zero in FSM calculations.

We included the orbital-dependent potentials, the so-called LDA$+ U$ (or GGA$+ U$) method.\cite{Anisimov:93} When we calculate the effective potential $U_{\textrm{eff}} = U - J$, we set the exchange parameter $J = 0$, and the onsite Coulomb interaction $U = - 0.09$ Ry and $- 0.10$ Ry for the 122- and 1111-compounds, respectively. These negative \emph{U}'s give very well-matched results with experimental ones. We used experimental lattice constants and magnetic structure. In the case of the 122-compound, the lattice constants were taken from Ref. \onlinecite{Rotter:08}, and magnetic structure was taken from Ref. \onlinecite{Huang:08}, and for the case of the 1111-compound, they were taken from Ref. \onlinecite{Nomura:08} and Ref. \onlinecite{Cruz:08}, respectively. From now on, we will refer to this method, \emph{i.e.} GGA$+U$ method with negative $U$, as GGA$-|U|$ method.

\section{Results and discussion}

The calculated magnetic moments obtained by a simple GGA calculations are about 2.02 $\mu_{\textrm{B}}$/Fe and 1.87 $\mu_{\textrm{B}}$/Fe for the  122- and 1111-compounds, respectively. These values are similar to those of previous publications ($\sim$2.6 $\mu_{\textrm{B}}$\cite{Akturk:09} and $\sim 1.6 \mu_{\textrm{B}}$\cite{Nomura:08} for the 122- and 1111-compounds, respectively), however, they are still much larger than those of experiments. They were reduced to be 0.87 $\mu_{\textrm{B}}$/Fe and 0.38 $\mu_{\textrm{B}}$/Fe obtained by using GGA$-|U|$ method with $U = - 0.09$ Ry and $- 0.10$ Ry, respectively, which are in excellent agreement with the experimental ones.\cite{Huang:08,Cruz:08}

To understand these changes, the differences between the symmetry-decomposed occupation numbers of \emph{d}-orbitals obtained by using the GGA and GGA$-|U|$ methods are summarized in Table \ref{occupation}. For the 122-compound the application of negative $U$ results in the reduction of the majority-spin \emph{d}-electron occupation number by $\sim 0.55$ and increase of the minority-spin \emph{d}-electron occupation number by $\sim 0.58$ resulting in a decrease of 1.13 $\mu_{\textrm{B}}$ of the total magnetic moment. The increase in minority-spin \emph{d}-electron occupation number is more significant in the $t_{2g}$ orbitals ($\sim 0.37$) than that of the $e_{g}$ orbitals ($\sim 0.21$). Almost the same number of electrons are decreased in majority-spin bands for the $t_{2g}$ and $e_{g}$ bands by 0.29 and 0.26 electrons, respectively. Since the $t_{2g}$ orbitals are composed of the $d_{xy}$, $d_{xz}$ and $d_{yz}$ orbitals, and the $e_{g}$ orbitals are composed of the $d_{(x2-y2)}$ and $d_{z^{2}}$ orbitals, the increased minority-spin \emph{d}-electron are almost evenly distributed to all 5 \emph{d}-orbitals, while the electrons per orbital in the majority-spin $e_{g}$ orbitals (0.13) are more severely decreased than those of the $t_{2g}$ orbitals (0.097). This difference may cause the symmetry changes of the spin-density plots, which will be discussed later. The similar arguments can be applied to the case of the 1111-compound.

\begin{table}
\caption{Differences between the symmetry-decomposed occupation numbers of \emph{d}-orbitals obtained by using GGA and GGA$-|U|$ methods ($\Delta {n_{{e_g} \uparrow }} \equiv n_{{e_g} \uparrow }^{{\rm{GGA}} - \left| U \right|} - n_{{e_g} \uparrow }^{{\rm{GGA}}}$, and so on). The change of symmetry-decomposed occupation numbers of \emph{d}-orbitals upon transition from high-spin to low-spin states in Fe$_{3}$Al alloy are also included for the reference.}\label{occupation}
\begin{ruledtabular}
\begin{tabular}{ccccc}
   & $\Delta n _{e_{g}\uparrow}$& $\Delta n _{e_{g}\downarrow}$& $\Delta n _{t_{2g}\uparrow}$& $\Delta n _{t_{2g}\downarrow}$\\
  \hline
  122 & $-$0.26 & 0.21 & $-$0.29 & 0.37\\
  1111 & $-$0.33 & 0.27 & $-$0.39 & 0.43\\
  Fe$_{3}$Al & $-$0.26 & 0.10 & $-$0.28 & 0.48 \\
\end{tabular}
\end{ruledtabular}
\end{table}

These redistributions of the symmetry-decomposed \emph{d}-orbital occupation numbers (\emph{d}-OONs) calculated by using the GGA$-|U|$ method is very similar to the case of those of Fe$_{3}$Al alloy upon transition from the high-spin to low-spin states.\cite{Rhee:04} As can be seen in the last row of Table \ref{occupation}, the reduction of the majority-spin \emph{d}-OON by $\sim$0.54, and the increase of the minority-spin \emph{d}-OON by $\sim$0.58, result in a total magnetic moment decrease of 1.22 $\mu_{\textrm{B}}$ in Fe$_{3}$Al alloy upon the transition from the high-spin to low-spin states. Therefore, we may think that the magnetic states of Fe atoms in the 122- and 1111-compounds are similar to the case of low-spin state of Fe$_{3}$Al alloy.

There are two crystallographically different atomic sites for Fe atoms in Fe$_{3}$Al alloy; \emph{i}) Fe$_{\textrm{I}}$ atoms and \emph{ii}) Fe$_{\textrm{II}}$ atoms. Fe$_{\textrm{I}}$ atoms are surrounded by 4 Fe$_{\textrm{II}}$ atoms and 4 Al atoms, and the surrounding atoms are located at the corners of a cube centered at the Fe$_{\textrm{I}}$ atom. On the other hand, Fe$_{\textrm{II}}$ atoms are surrounded by 8 Fe$_{\textrm{I}}$ atoms. The Fe$_{\textrm{I}}$ atom exhibits the transition from the high-spin to low-spin states at high pressure, while the magnetic moment of Fe$_{\textrm{II}}$ atom remains almost unchanged.

The 4 Al atoms surrounding a Fe$_{\textrm{I}}$ atom form a tetrahedron. It is very similar to the tetrahedral cages formed by As ligand atoms in the 122- and 1111-compounds. According to the crystal-field theory,\cite{wiki:CFT} the 5 \emph{d}-orbitals will be split into the $t_{2g}$ and $e_{g}$ orbitals if the transition metal is surrounded by 4 ligand atoms forming a tetrahedral cage. Since the ligand atoms of tetrahedral cage are directly contacted with one of the lobes of $t_{2g}$ orbitals, these orbitals have high electron-electron repulsion, resulting in a higher energy than the $e_{g}$ orbitals. Since the amount of direct contact is not very large, the energy splitting is not very significant compared with the case of octahedral cages. If the energy difference ($\Delta E_{t}$) between $t_{2g}$ and $e_{g}$ orbitals is smaller than the pairing energy ($\Delta E_{p}$), then two electrons with opposite spins prefer to form a electron pair because the pairing reduces the total energy and, thus, the electrons do not obey the Hund's rule. The pairing reduces the magnetic moments and the transition metal-ligands complex (TMLC) is in the low-spin state. In the majority of cases, however, $\Delta E_{t}<\Delta E_{p}$. Therefore, most of tetrahedral TMLCs are in the high-spin state.

The transition from high-spin state to low-spin state at high pressure can be understood by the fact that a decreased lattice constant results in pushing the bands above (below) the Fermi level ($E_{\textrm{F}}$) toward the higher (lower) energy region.\cite{Rhee:00,Rhee:03} It implies that $\Delta E_{t}$ increases as the pressure is applied, favoring the pairing of electrons and, hence, the transition to the low-spin state

\begin{figure}[tbp]
\begin{center}
\includegraphics[width=9cm]{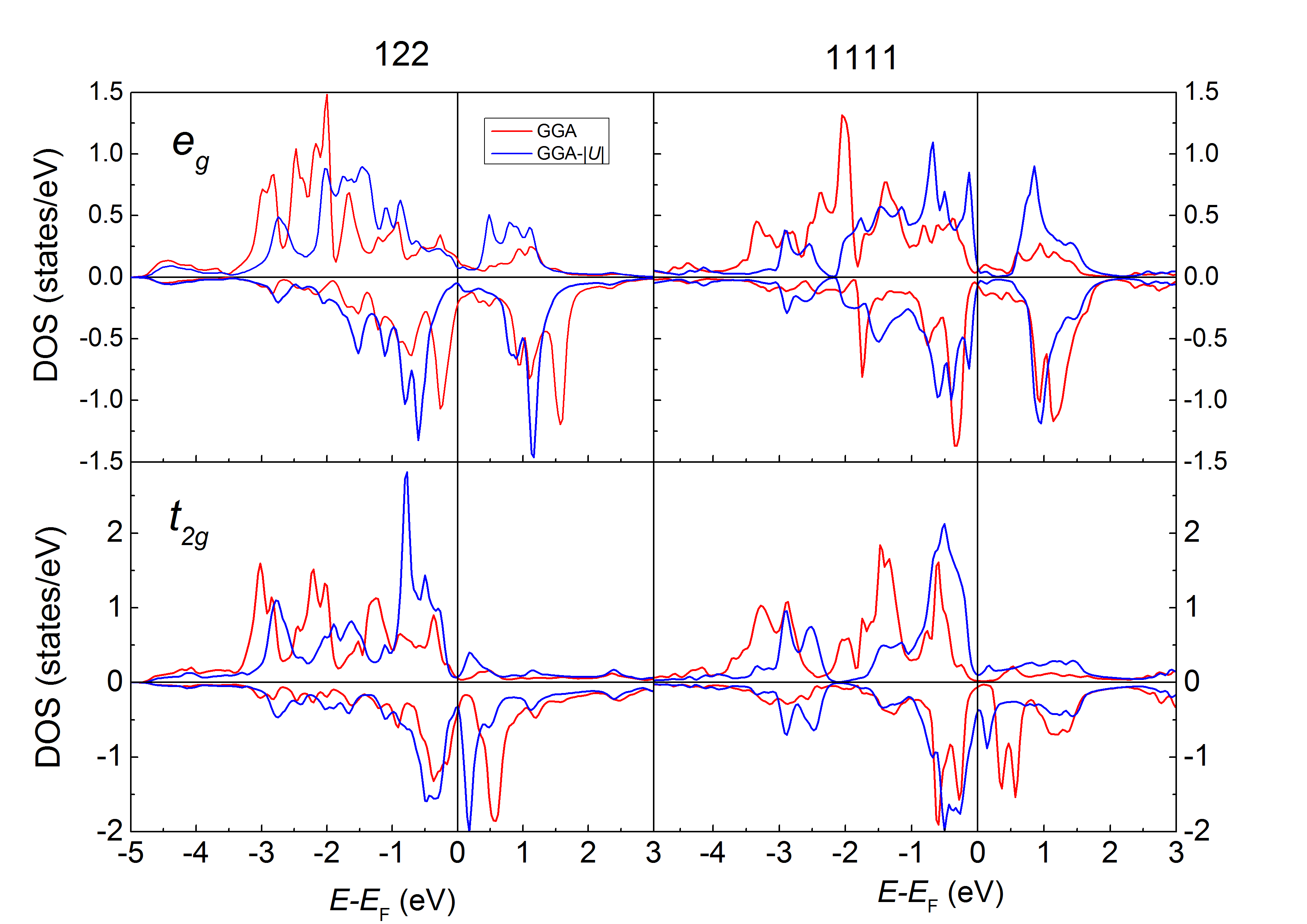}
\caption{(Color online.) Symmetry-decomposed Fe \emph{d}-DOS of the 122- and 1111-compounds.}
\label{DOS}
\end{center}
\end{figure}

\begin{figure}[tbp]
\begin{center}
\includegraphics[width=8cm]{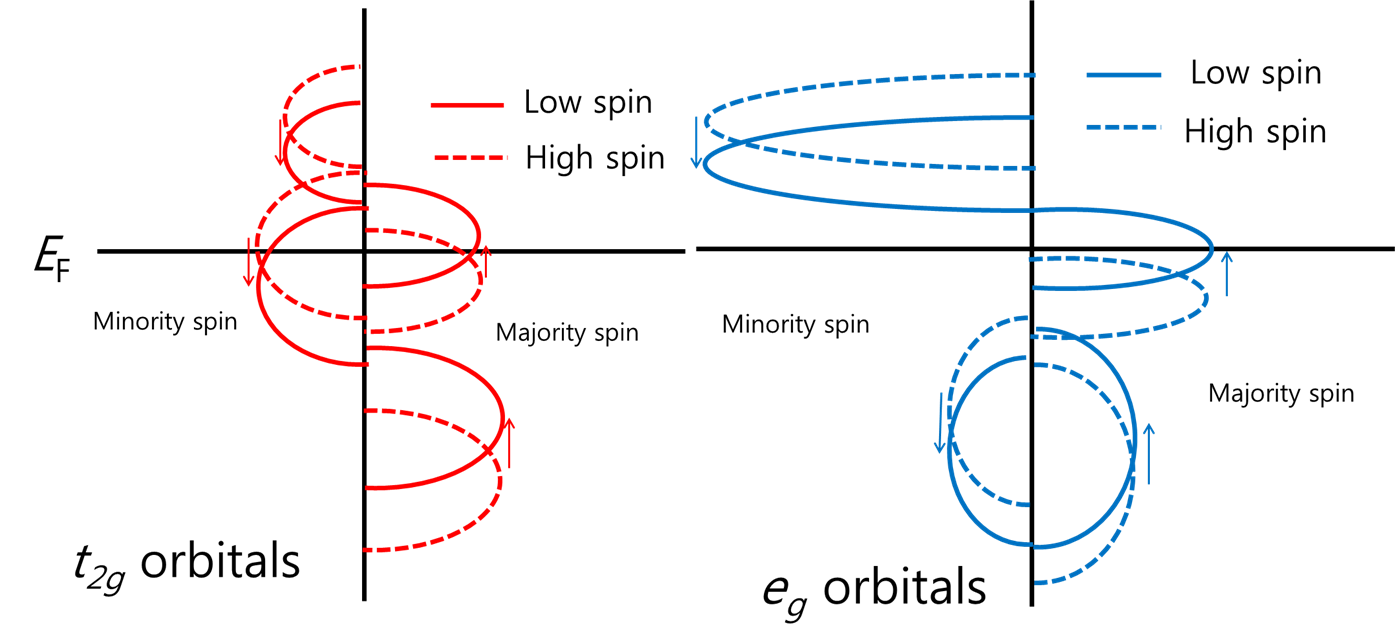}
\caption{(Color online.) Schematic diagrams of symmetry-decomposed Fe \emph{d}-DOS of metamagnetic Fe$_{3}$Al alloy.}
\label{DOS_schematic}
\end{center}
\end{figure}

In our case, the situation is not that simple. Figure \ref{DOS} clearly exhibits the tendency of redistribution of the symmetry-decomposed \emph{d}-OONs of both compounds. Majority-spin bands move upward, while minority-spin bands move downward for both $t_{2g}$ and $e_{g}$ orbitals, and all $t_{2g}$ and $e_{g}$ orbitals are split into two band groups for both majority- and minority-spin bands. To show this tendency of redistribution of the symmetry-decomposed \emph{d}-OONs more clearly a schematic diagrams of symmetry decomposed Fe \emph{d}-DOS of metamagnetic Fe$_{3}$Al alloy is presented in Fig \ref{DOS_schematic}. For the $e_{g}$ majority-spin bands only the upper-band group is located near $E_{\textrm{F}}$, while the both band groups of minority-spin bands are far from $E_{\textrm{F}}$. Therefore, \emph{d}-OONs of $e_{g}$ orbitals do not change significantly for majority spin, while there are almost no change in \emph{d}-OONs of $e_{g}$ orbitals for minority spin upon transition to the low-spin state. The situation of the case of $t_{2g}$ orbitals is quite different from that of the $e_{g}$ orbitals. The upper majority-spin bands of $t_{2g}$ orbitals are located just below $E_{\textrm{F}}$, and the lower minority-spin bands of $t_{2g}$ orbitals are located just above $E_{\textrm{F}}$ for the high-spin state. Upon transition to the low-spin state, \emph{d}-OONs for majority-spin (minority-spin) $t_{2g}$ orbitals decreases (increases) significantly, resulting in a significant reduction of magnetic moment.

\begin{figure}[tbp]
\begin{center}
\includegraphics[width=8.5cm]{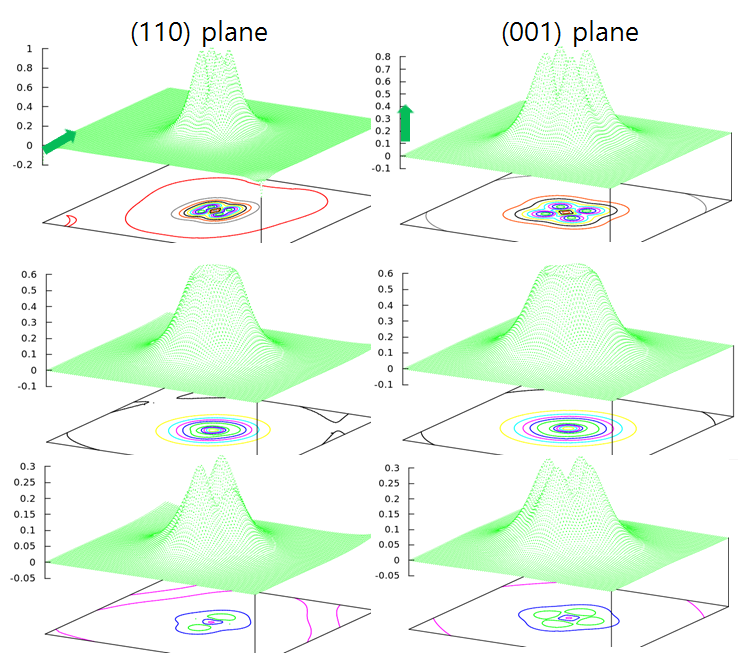}
\caption{(Color online.) Spin-density plots of the Fe$_{3}$Al alloy on the (110) and (001) planes for (top) fictitious Fe$_{2}$Al alloy (see text), (middle) for high-spin state, and (bottom) for the low-spin state. Thick arrows indicate the direction of magnetization.}
\label{Fig_HS-LS-Fe3Al}
\end{center}
\end{figure}

This redistributions of \emph{d}-OONs can be clearly manifested in spin-density plots. Before discussing the spin densities of the Fe-based pnictide compounds, we present spin-density plots for the high-spin and low-spin states of metamagnetic Fe$_{3}$Al alloy in Fig. \ref{Fig_HS-LS-Fe3Al}. The distribution of spin density plotted on the (110) plane (left panel) around the Fe$_{\textrm{I}}$ atom is symmetric and exhibits no significant directionality in the high-spin state, while that for the low-spin state exhibits a strong directionality. Especially, the strong directionality appears along the \emph{z}-axis. For the high-spin state \emph{d}-OONs of five \emph{d}-orbitals are almost identical to each other, while those of $e_{g}$ orbitals are significantly smaller than those of $t_{2g}$ orbitals for the low-spin state. Therefore, the directionality appeared along the \emph{z}-axis is evident, reflecting the deficiency of $d_{z^{2}}$ orbital in the low-spin state. The same arguments can be applied to the plots on the (001)-plane (see the right panel of Fig. \ref{Fig_HS-LS-Fe3Al}). In the low-spin state the spin density exhibits strong directionality along the \emph{x}- and \emph{y}-axes, reflecting the deficiency of $d_{(x^{2}-y^{2})}$ orbitals.

\begin{figure}[tbp]
\begin{center}
\includegraphics[width=8.5cm]{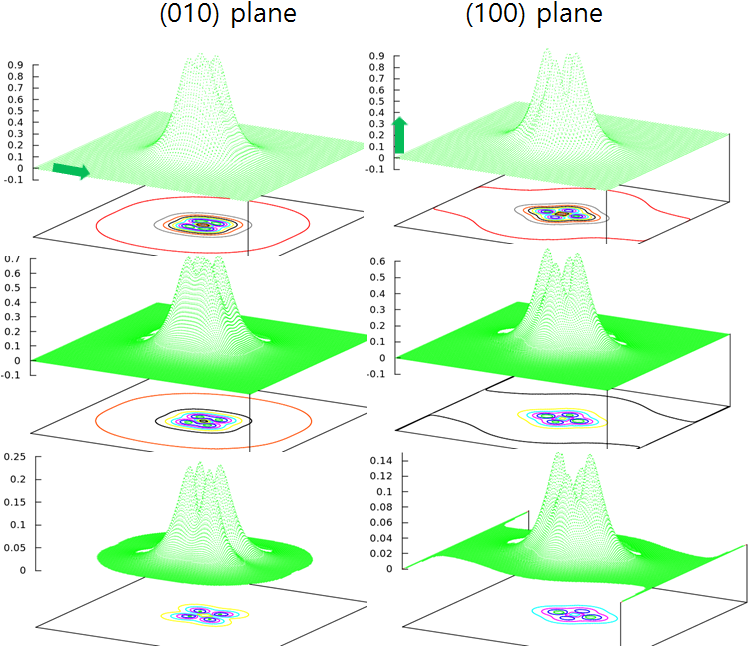}
\caption{(Color online.) Spin-density plots of the 1111-compound on the (010) and (100) planes for (top) fictitious compound with expanded lattice constant (see text), (middle) for  a simple GGA calculation, and (bottom) for GGA$-|U|$  calculation. Thick arrows indicate the direction of magnetization.}
\label{Fig_SD_1111}
\end{center}
\end{figure}

To see if the same trend can be found in the Fe-based pnictide compounds, Fig. \ref{Fig_SD_1111} presents the spin-density plots of the 1111-compound on the (010) and (100) planes. Unlike the case of the metamagnetic Fe$_{3}$Al alloy, in calculations with simple GGA there is no symmetric distribution of spin density on the (010) plane, in which the magnetization is in the plane. In the GGA$-|U|$ calculation the spin density exhibits clear 4-fold symmetry, while it is slightly 2-fold in the simple GGA calculation. For the (100) plane, in which the magnetization is out of the plane, the situation becomes opposite. It exhibits almost perfect 4-fold symmetry in the simple GGA calculation, while it becomes slightly 2-fold symmetric in the the GGA$-|U|$ calculation. Very similar arguments can be applied to the case of the 122-compound.

Since the transition from high-spin state to low-spin state occurs at high pressure, the distance between the Fe atom and ligand atom, ${d_{{\rm{Fe - Lig.}}}}$, may play a crucial role. For the Fe$_{3}$Al alloy ${d_{{\rm{Fe - Lig.}}}}= 4.48$ a.u.\cite{Rhee:04} in the low-spin state, while those for the 122- and 1111-compounds are 4.53 a.u. and 4.46 a.u., respectively. Since all ${d_{{\rm{Fe - Lig.}}}}$'s are in almost the same distances, we calculated the spin densities of a fictitious 1111-compound with the expanded lattice constant, bearing ${d_{{\rm{Fe - Lig.}}}}= 4.7$ a.u., which is the same as that of Fe$_{3}$Al alloy in the ambient pressure. As can be seen in the top row of Fig. \ref{Fig_SD_1111}, the spin-density plots do not exhibit a symmetric distribution of spin density. Rather, it is more 2-fold symmetric for the (010) plane and more 4-fold symmetric for the (100) plane. Although there are only 4 As atoms, forming a tetrahedral cage, around the Fe atoms in Fe-based pnictide compounds, there are 4 Al atoms, forming a tetrahedral cage, \emph{plus} 4 Fe$_{\textrm{II}}$ atoms, also forming a tetrahedral cage, around the Fe$_{\textrm{I}}$ atom. To compare the results of the Fe$_{3}$Al alloy with those of Fe-based pnictide compounds, it is necessary to remove the effects of surrounding 4 Fe$_{\textrm{II}}$ atoms. Therefore, we calculated the spin density of another fictitious Fe$_{2}$Al alloy, in which the Fe$_{\textrm{II}}$ atoms are removed from the Fe$_{3}$Al alloy. The results are almost identical to those of the fictitious 1111-compound with the expanded lattice constant (see the top row of Fig. \ref{Fig_HS-LS-Fe3Al}).

The symmetric distribution of spin density in the high-spin state of the Fe$_{3}$Al alloy can be attributed to the existence of 4 Fe$_{\textrm{II}}$ atoms. Since the 4 Fe$_{\textrm{II}}$ atoms form another tetrahedral cage, they can attract spin density into themselves from the Fe$_{\textrm{I}}$ atom, resulting in a symmetric distribution of spin density in the high-spin state. The effects of the  existence of 4 Fe$_{\textrm{II}}$ atoms can be further manifested by the fact that the symmetry directions in the spin-density plot on the (001) plane for the low-spin state are 45$^{\circ}$ rotated from those of fictitious Fe$_{2}$Al alloy.

According to our calculational results, the LDA, GGA or GGA$+U$ with positive \emph{U} can not properly predict the magnetic states of the Fe-based pnictide superconductors. Rather, the GGA$-|U|$ method can. In Ref. \onlinecite{Nakamura:09} it was argued that the negative \emph{U} correction can be considered as an unexpectedly well screening on \emph{d}-orbitals in Fe atoms. If the screening is strong enough or over, then we can reach a situation that the intra-band repulsion becomes smaller than the inter-band one. In such a case, an effectively attractive force may result in.

\section{Summary}
We have calculated the electronic structures, DOS, occupation numbers and spin-density of 3\emph{d} Fe of BaFe$_{2}$As$_{2}$ and LaFeAsO in the orthorhombic, AFM phases by using GGA and GGA$-|U|$. We found that, in high-$T{_\textrm{c}}$ iron-based superconducting compounds, the use of GGA with negative \emph{U} with similar values ($- 0.09$ Ry and $- 0.10$ Ry) give the well-matched results with experimental magnetic moment per Fe atom. By comparing the differences of DOS, symmetry-decomposed \emph{d}-OONs and spin densities between simple GGA and GGA-$|U|$ calulations, the magnetic state of the iron-based pnictide compounds are very similar to the case of the low-spin state of metamagnetic Fe$_{3}$Al alloy. To further address the validity of negative \emph{U} more researches, such as direct calculation of the onsite Coulomb interaction \emph{U} and the exchange iteration \emph{J} from the first-principles calculations, are needed.

\acknowledgments This research was supported by the Basic Science Research Program through the National Research Foundation of Korea (NRF) funded by the Ministry of Education, Science and Technology (2011-0023423). This work was also supported by the Faculty Research Fund, Sungkyunkwan University, 2006.
\bibliography{FePnictide}

\end{document}